\documentclass[aps,pre,reprint,groupedaddress]{revtex4-2}

\usepackage{amsmath}
\usepackage{amssymb}
\usepackage{graphicx}
\usepackage{xcolor}
\usepackage[normalem]{ulem}

\begin{document}


\title{Structure of fluctuating thin sheets under random forcing}


\author{Chanania Steinbock}
\author{Eytan Katzav}
\affiliation{Racah Institute of Physics, The Hebrew University, Jerusalem 9190401, Israel}
\author{Arezki Boudaoud}
\affiliation{LadHyX, CNRS, Ecole Polytechnique, Institut Polytechnique de Paris, Palaiseau Cedex 91128 France}


\date{\today}

\begin{abstract}
We propose a mathematical model to describe the athermal fluctuations of thin sheets driven by the type of random driving that might be experienced prior to weak crumpling. The model is obtained by merging the F\"oppl-von K\'arm\'an equations from elasticity theory with techniques from out-of-equilibrium statistical physics to obtain a nonlinear strongly coupled $\phi^{4}$-Langevin field equation with spatially varying kernel. With the aid of the self-consistent expansion (SCE), this equation is analytically solved for the structure factor of a fluctuating sheet. In contrast to previous research which has suggested that the structure factor follows an anomalous power-law, we find that the structure factor  in fact obeys a logarithmically corrected rational function. Numerical simulations of our model confirm the accuracy of our analytical solution.
\end{abstract}


\maketitle

\section{Introduction\label{sec:intro}}

Thin sheets are able to exhibit a remarkable variety of structures with extreme variation in complexity. For example, the same sheet of paper can be used to construct a simple greeting card, an intricate work of origami or a ``mere'' crumpled ball. Though obtaining the crumpled ball certainly requires less deliberation, the underlying structure is no less complex than that of the origami artwork. Indeed, though the structure of a randomly crumpled sheet has been the subject of active research, both experimentally \cite{PlouraboueRoux1996, Matan2002, BlairKudrolli2005, Balankin2006, Andresen2007, Balankin2008, Deboeuf2013, Balankin2013, Gottesman2018} and through modelling \cite{PlouraboueRoux1996, Vliegenthart2006, Sultan2006, Andrejevic2021}, a mathematical theory derived from first principles remains lacking.

Early work on crumpling focused on membranes or so called ``tethered sheets'' in which a two-dimensional triangular lattice of spheres with fixed distances between neighbouring spheres is randomly fluctuated in space \cite{Kantor1986, Kantor1987, Kantor1987a, Kardar1987, Paczuski1988, Duplantier1990}. The great advantage of such models is that they are readily generalised to include the effects of self-avoidance, an important property in the context of strong crumpling \cite{Vliegenthart2006}. An interesting insight that was used in these early works is a useful analogy with two-dimensional models of Heisenberg ferromagnetism, where one could think of the normals to the surface as Heisenberg spins. In this framework, a flat sheet corresponds to a ferromagnetic state while a crumpled sheet to a paramagnetic state.

The subsequent development of a mathematical theory which included elastic effects led to a description of the various types of singular structures which can arise within thin sheets, notably,  structures such as ridges and d-cones \citep{Lobkovsky1995, Lobkovsky1997, BenAmar1997, Chaieb1997, Cerda1999, Mora2002, Liang2005}. From the theory of elasticity, it is known that thin sheets undergo two types of deformation -- energetically cheap bending and energetically expensive stretching. The theory of ridges and d-cones describes from first principles how stretching becomes highly localised in thin sheets to minimise the energy expenditure and these become the creases of a folded sheet. This theory, however, has only been successfully applied to characterise situations with relatively few ridges. 

In this paper, we propose a different approach, combining methods from elasticity theory with out-of-equilibrium statistical physics. From elasticity theory, we consider the F\"oppl-von K\'arm\'an equations -- the theory of linear elasticity applied to thin sheets while also accounting for the geometric nonlinear effects of deformation \cite{LandauLifshitzElasticityBook}. The result is a pair of coupled nonlinear partial differential equations which describe the deformation of thin sheets and though these equations are difficult to solve in general, it has indeed been shown that ridges and d-cones are particular solutions \cite{Lobkovsky1997, BenAmar1997}. By studying a dynamic noisy variant of these equations, we are able to model a thin sheet subjected to a continually varying athermal random force. In this manner, we are able to begin characterising the structure of a randomly fluctuating thin sheet, a structure which could conceivably resemble to some degree the structure of a crumpled sheet. The underlying logic here being that it is the pure elastic deformations that set the fundamental modes of the sheet and these are then plastified into creases by some irreversible process when the curvature is large. From this perspective, a weakly crumpled sheet is a nonlinear echo of its elastic fluctuations. Similar arguments are indeed often invoked in the study of irreversible processes in mechanical systems. A famous example being the fragmentation of thin rods, described in \cite{Gladden2005, Audoly2005}, where the points that a buckled thin rod snaps are the points of maximal curvature of the purely elastically deformed rod.

Indeed, the question of characterising the structure of fluctuating thin sheets is a fundamental one in its own right with ramifications spanning a diversity of fields \cite{NelsonBook2004}. Applications range from the biophysics of cell membranes \cite{Liang2016}, to the properties of graphene \cite{Meyer2007, Meyer2007a, Thompson2009, Zang2013, Deng2016, Ahmadpoor2017}, to the character of wave turbulence \cite{During2006, Boudaoud2008, Mordant2008, Cadot2008, Cobelli2009, Humbert2013, During2015, During2017, During2019}. In recent years, it has become apparent that the precise structure of fluctuating thin sheets is important for understanding their mechanical and elastic properties \cite{Kosmrlj2013, Kosmrlj2014, Kosmrlj2016, Kosmrlj2017, Shankar2021, Morshedifard2021} as well as their acoustic \cite{Kramer1996, Houle1996} and optical emissions \cite{Rad2019}. Accordingly, we consider the lack of an out-of-equilibrium model for fluctuating surfaces as a deficiency which we begin to remedy here.

The paper is organised as follows. In Section \ref{sec:derivation}, we outline and prepare the variant of the F\"oppl-von K\'arm\'an equations which we intend to study while in Section \ref{sec:SCE}, we use a method known as the self-consistent expansion (SCE) to obtain analytical solutions for the structure factor of a fluctuating sheet. In Section \ref{sec:simulations}, our solution is compared with numerical simulations of the dynamic F\"oppl-von K\'arm\'an equations and the conclusions are discussed in Section \ref{sec:discussion}. Various technical details of the solution derived in Section \ref{sec:SCE} are postponed to an Appendix.

\section{The Overdamped Dynamic F\"oppl-von K\'arm\'an Equations\label{sec:derivation}}

The F\"oppl-von K\'arm\'an equations
\begin{align}
P_{ex} &= \frac{Eh^{3}}{12\left(1-\nu^{2}\right)}\nabla^{4}\xi \nonumber\\
&-h\left(\frac{\partial^2\xi}{\partial x^{2}}\frac{\partial^{2}\chi}{\partial y^{2}}+\frac{\partial^{2}\xi}{\partial y^{2}}\frac{\partial^{2}\chi}{\partial x^{2}}-2\frac{\partial^{2}\xi}{\partial x\partial y}\frac{\partial^{2}\chi}{\partial x\partial y}\right)\\
0 &= \nabla^{4}\chi+E\left[\frac{\partial^{2}\xi}{\partial x^{2}}\frac{\partial^{2}\xi}{\partial y^{2}}-\left(\frac{\partial^{2}\xi}{\partial x\partial y}\right)^{2}\right]\label{eq:fvk airy}
\end{align}
have long been used to study the equilibrium out-of-plane displacement $\xi\left(x,y\right)$ of a thin sheet subject to an external pressure $P_{ex}$. Here, $h$ denotes the thickness of the sheet, $E$ its Young's modulus and $\nu$ its Poisson ratio. The scalar field $\chi\left(x,y\right)$ is the Airy stress potential of the deformation. The Monge parameterisation used, in which the vertical displacement $\xi(x,y)$ is written as a function of $x$ and $y$, implies that we consider deformations that are mostly flat and thus we will only focus in the continuation on weak fluctuations. Clearly, self-avoidance of any real sheet further strengthens this assumption.

To study a driven system, we simply apply Newton's second law to each element of the sheet with density $\rho$
\begin{equation}
h\rho\frac{\partial^{2}\xi}{\partial t^{2}}=-P_{ex}+P_{damping}+P_{driving}\,.
\end{equation}
Though similar equations have been studied in the context of wave turbulence \cite{During2006, Boudaoud2008, Mordant2008, Cadot2008, Cobelli2009, Humbert2013, During2015, During2017, During2019}, our approach deviates from this prior research in that we consider specific damping and driving pressures. In particular, for the damping, we consider ordinary fluid friction $P_{damping}=-\alpha\frac{\partial\xi}{\partial t}$ where $\alpha$ denotes the coefficient of friction while for the driving, we consider zero-mean conserved Gaussian noise $P_{driving}=\eta\left(\vec{r},t\right)$ with noise amplitude $D$, that is,
\begin{align}
\left\langle \eta\left(\vec{r},t\right)\right\rangle &=0\\
\left\langle \eta\left(\vec{r},t\right)\eta\left(\vec{r}\,',t'\right)\right\rangle &=-D\delta\left(t-t'\right)\nabla^{2}\delta\left(\vec{r}-\vec{r}\,'\right)\,.
\end{align}
While other driving forces such as thermal fluctuations could certainly be considered, our aim here is to develop a minimal model for the type of athermal fluctuations one might expect immediately prior to crumpling. In the spirit of critical phenomena models that incorporate the symmetries and conserved quantities of their systems, we impose conserved noise on the sheet as this ensures that the sheet's center of mass will not wander in space. Indeed, any more specific form of noise which is consistent with the conservation of center of mass could also be considered. Applying the principle of parsimony however, any more specific form would need additional justification, for example, by being experimentally motivated.

Finally, in further contrast to the wave turbulence context, we focus on the overdamped limit in which the inertia term $h\rho\frac{\partial^{2}\xi}{\partial t^{2}}$ can be neglected relative to the damping term. Ultimately, this gives the overdamped dynamic F\"oppl-von K\'arm\'an equation
\begin{multline}
\alpha\frac{\partial\xi}{\partial t}+\frac{Eh^{3}}{12\left(1-\nu^{2}\right)}\nabla^{4}\xi\\
-h\left(\frac{\partial^2\xi}{\partial x^{2}}\frac{\partial^{2}\chi}{\partial y^{2}}+\frac{\partial^{2}\xi}{\partial y^{2}}\frac{\partial^{2}\chi}{\partial x^{2}}-2\frac{\partial^{2}\xi}{\partial x\partial y}\frac{\partial^{2}\chi}{\partial x\partial y}\right)=\eta\left(\vec{r},t\right)\,.\label{eq:fvk dynamic}
\end{multline}
Together with equation (\ref{eq:fvk airy}), this equation describes the stochastic time varying deformation of a sheet fluctuating under the influence of conserved Gaussian noise. Alternatively, for a sheet of dimensions $L\times L$, equations (\ref{eq:fvk airy}) and (\ref{eq:fvk dynamic}) can be written in Fourier space and combined into the single equation
\begin{multline}
\alpha\frac{\partial\tilde{\xi}_{\vec{n}}}{\partial t}+\frac{\left(2\pi\right)^{4}}{L^{4}}\frac{Eh^{3}}{12\left(1-\nu^{2}\right)}\left|\vec{n}\right|^{4}\tilde{\xi}_{\vec{n}}\\
+\frac{\left(2\pi\right)^{4}}{L^{4}}\frac{hE}{2}\sum_{\vec{\ell}_{1}\ne\vec{n}}\sum_{\vec{\ell}_{2}}\sum_{\vec{\ell}_{3}}V_{\vec{n},\vec{\ell}_{1},\vec{\ell}_{2},\vec{\ell}_{3}}\tilde{\xi}_{\vec{\ell}_{1}}\tilde{\xi}_{\vec{\ell}_{2}}\tilde{\xi}_{\vec{\ell}_{3}}=\tilde{\eta}_{\vec{n}}\left(t\right)\,,\label{eq:fvk fourier}
\end{multline}
where the kernel of the sum $ V_{\vec{n},\vec{\ell}_{1},\vec{\ell}_{2},\vec{\ell}_{3}} $, given by
\begin{equation}
V_{\vec{n},\vec{\ell}_{1},\vec{\ell}_{2},\vec{\ell}_{3}}=\delta_{\vec{n},\vec{\ell}_{1}+\vec{\ell}_{2}+\vec{\ell}_{3}}\frac{\left|\vec{n}\times\vec{\ell}_{1}\right|^{2}\left|\vec{\ell}_{2}\times\vec{\ell}_{3}\right|^{2}}{\left|\vec{n}-\vec{\ell}_{1}\right|^{4}}\,,
\end{equation}
is simply the Fourier transform of the transverse projection operator of the sheet deformation \cite{NelsonBook2004, Nelson1987}, having denoted $\vec{n}\times\vec{\ell}=n_{x}\ell_{y}-n_{y}\ell_{x}$. Here, $\tilde{\xi}_{\vec{n}} = \tilde{\xi}_{\vec{n}}(t)$ and $\tilde{\eta}_{\vec{n}}\left(t\right)$ denote the Fourier components of $\xi\left(\vec{r},t\right)$ and $\eta\left(\vec{r},t\right)$ respectively, that is $\xi\left(\vec{r},t\right)=\sum_{\vec{n}}\tilde{\xi}_{\vec{n}}(t)e^{i\frac{2\pi}{L}\vec{n}\cdot\vec{r}}$ and similarly for $\eta\left(\vec{r},t\right)$. The sums in equation (\ref{eq:fvk fourier}) are carried out over all integer lattice points of $\mathbb{R}^{2}$ excluding the point $\vec{\ell}_{1}=\vec{n}$. As the nonlinear term of equation (\ref{eq:fvk fourier}) is essentially a quartic interaction, this equation can also be thought of as a type of Langevin $\phi^{4}$-field theory \cite{Kleinert2001} decorated by a non-trivial spatially varying kernel $V_{\vec{n},\vec{\ell}_{1},\vec{\ell}_{2},\vec{\ell}_{3}}$. It is worth noting that the zeroth mode, $\vec{n}=0$, of equation (\ref{eq:fvk fourier}) vanishes and thus $\tilde{\xi}_{\vec{0}} $ remains constant in time. Accordingly, not only is the noise conservative (by design) but the deterministic forces are as well, though this conservation is achieved in a different manner than model~B within the classical classification of Hohenberg and Halperin \cite{Hohenberg1977} where a Laplacian is also applied to the deterministic forces.

It will be convenient to work with a nondimensionalised form of equation (\ref{eq:fvk fourier}). To this end, we can define the dimensionless length, time, vertical displacement and noise in Fourier space
\begin{align}
\bar{r} &=\vec{r}/L\\
\bar{t}&=\left[\left(2\pi\right)^{6}hDE/\left(\alpha^{3}L^{8}\right)\right]^{1/2}t\\
\bar{\xi}_{\vec{n}}(\bar{t}) &=\left[\left(2\pi\right)^{2}\alpha hE/D\right]^{1/4}\tilde{\xi}_{\vec{n}}(t)\\
\bar{\eta}_{\vec{n}}\left(\bar{t}\right) &=\left[\alpha^{3}L^{16}/\left(\left(2\pi\right)^{10}hD^{3}E\right)\right]^{1/4}\tilde{\eta}_{\vec{n}}\left(t\right)
\end{align}
which gives rise to the nondimensionalised equation
\begin{multline}
\frac{\partial\bar{\xi}_{\vec{n}}}{\partial\bar{t}}+g\left|\vec{n}\right|^{4}\bar{\xi}_{\vec{n}}\\
+\frac{1}{2}\sum_{\vec{\ell}_{1}\ne\vec{n}}\sum_{\vec{\ell}_{2}}\sum_{\vec{\ell}_{3}}V_{\vec{n},\vec{\ell}_{1},\vec{\ell}_{2},\vec{\ell}_{3}}\bar{\xi}_{\vec{\ell}_{1}}\bar{\xi}_{\vec{\ell}_{2}}\bar{\xi}_{\vec{\ell}_{3}}=\bar{\eta}_{\vec{n}}\left(\bar{t}\,\right)\label{eq:fvk fourier dmnless}
\end{multline}
containing the single dimensionless parameter
\begin{equation}
g=\frac{2\pi}{12\left(1-\nu^{2}\right)}\sqrt{\frac{\alpha h^{5}E}{D}}
\end{equation}
and nondimensionalised zero-mean conserved Gaussian noise with covariance
\begin{equation}
\left\langle \bar{\eta}_{\vec{n}}\left(\bar{t}\,\right)\bar{\eta}_{\vec{n}'}\left(\bar{t}\,'\right)\right\rangle =\left|\vec{n}\right|^{2}\delta_{\vec{n},-\vec{n}'}\delta\left(\bar{t}-\bar{t}\,'\right)\,.
\end{equation}

Since the dimensionless parameter $g$ scales like $h^{5/2}$, $g$ will be small for sufficiently thin sheets. Indeed, a typical sheet of aluminium foil or steel would have a thickness $h\sim0.5\mathrm{\,mm}$ and Young's modulus $E\sim200\mathrm{\,GPa}$. The magnitude of the driving force needed to induce fluctuations has been measured in \cite{Cadot2008} as being around $D\sim5\times10^{-3}\mathrm{\,N^{2}/Hz}$. The frequency dependent frictional damping rate $\gamma(f)$ of a fluctuating steel sheet has been measured in \cite{Humbert2013} as being constant for small frequencies and slowly growing for large frequencies at a rate of $\gamma \sim f^{0.6}$. Since we intend to describe the large scale features of a fluctuating sheet and $ \gamma $ anyway grows slowly with large frequencies, little is lost by treating $ \gamma $ as constant. From \cite{Humbert2013}, we thus take $\gamma\sim5\mathrm{\,Hz}$. Since the coefficient of friction is simply $\alpha=\rho h\gamma$ where $\rho\sim8\times10^{3}\mathrm{\,kg\cdot m^{-3}}$ denotes the density of the sheet and $\left(1-\nu^{2}\right)\sim1$, we obtain under typical circumstances that $g\sim0.1$ which is indeed significantly smaller than unity. For even thinner sheets or more violent driving forces, $g$ will only shrink and thus it is clear that the physically interesting case is indeed $g\ll1$.

Equation (\ref{eq:fvk fourier dmnless}) is a nonlinear Langevin equation for a type of quartic interaction and thus describes the sheet's structure both in-equilibrium and out-of-equilibrium, that is, in principle, equation (\ref{eq:fvk fourier dmnless}) can be used to obtain both static moments (equal time) such as the sheet's structure factor 
\begin{equation}
S_{\vec{n}}=\left\langle \bar{\xi}_{\vec{n}}(\bar{t}\,)\bar{\xi}_{-\vec{n}}(\bar{t}\,)\right\rangle \,
\end{equation}
and time-dependent moments such as the time-dependent correlation function
\begin{equation}
S_{\vec{n}}(\bar{t}_1,\bar{t}_2)=\left\langle \bar{\xi}_{\vec{n}}(\bar{t}_1)\bar{\xi}_{-\vec{n}}(\bar{t}_2)\right\rangle \,.
\end{equation}
Though the methods used in the continuation can be applied to both static and time-dependent quantities, in the following, we will focus only on determining the static structure factor and will defer the slightly more involved analysis of time-dependent quantities to the future.

It is interesting to mention here that based on the analogy to the Heisenberg model, the static structure factor is also related to the correlations between normals to the surface. As explained in \cite{Nelson1987} the tipping angles $\theta(x,y)$ of the normals relative to the $z$ axis obey approximately $\langle \theta^2(x,y) \rangle \simeq \langle |\nabla \xi (x,y)|^2 \rangle$.

Examining equation (\ref{eq:fvk fourier dmnless}), one sees that the small parameter $g$ appears in front of the linear part and thus we find that the fluctuations in the vertical displacement are strongly coupled to the nonlinear term. On the other hand, the linear term also grows rapidly with $\vec{n}$ since it is proportional to $\left| \vec{n} \right|^4$. As such, any perturbative expansion around the linear part of equation (\ref{eq:fvk fourier dmnless}) which merely treats the non-linear term as a small correction would be meaningless. At the same time, the linear term itself can also not simply be neglected. Accordingly, a more sophisticated approach is required.

\section{The Self-Consistent Expansion (SCE)\label{sec:SCE}}

The self-consistent expansion (SCE) can be conceived of as a renormalised perturbation theory \cite{McComb2003} capable of providing series approximations even in the presence of strong coupling. It has been used to solve systems similar to our own such as the KPZ equation and its variations \cite{Schwartz1992, Schwartz1998, Katzav1999, Katzav2002, Schwartz2002, Katzav2002a, Katzav2003, Katzav2004, Katzav2004a}, fracture and wetting fronts \cite{Katzav2006, Katzav2007} and turbulence \cite{Edwards2002}. Additionally, the SCE is known to provide an extremely successful solution to the zero-dimensional $\phi^{4}$-theory giving good results at low orders and providing exact convergence at high orders \cite{Schwartz2008, Remez2018}. Accordingly, it is reasonable to suppose that the SCE might shed light on our system as well. Though both simple self-consistent arguments \cite{Nelson1987} and more sophisticated self-consistent epsilon expansions in $d = 4 - \varepsilon$ dimensions \cite{LeDoussal1992, LeDoussal2018} have previously been applied to closely related problems, the approach presented here has the advantages of being natural, transparent and directly applicable to the problem at hand.

\subsection{The Fokker-Planck Equation}

To apply the SCE to our system, it is convenient to write its corresponding Fokker-Planck equation \cite{RiskenBook}
\begin{equation}
\frac{\partial P}{\partial\bar{t}}=\mathcal{O}P
\label{eq:FP operator}
\end{equation}
where $P=P\left(\left\{ \bar{\xi}_{\vec{n}}\right\} ,\bar{t}\,\right)$ denotes the probability that the system will have Fourier components $\left\{ \bar{\xi}_{\vec{n}}\right\}$  at time $\bar{t}$ and $\mathcal{O}$ denotes the Fokker-Planck operator for our system. For equation (\ref{eq:fvk fourier dmnless}), the Fokker-Planck equation takes the explicit form
\begin{multline}
\frac{\partial P}{\partial\bar{t}}=\frac{1}{2}\sum_{\vec{n}}\left|\vec{n}\right|^{2}\frac{\partial^{2}P}{\partial\bar{\xi}_{\vec{n}}\partial\bar{\xi}_{-\vec{n}}}+g\sum_{\vec{n}}\left|\vec{n}\right|^{4}\frac{\partial}{\partial\bar{\xi}_{\vec{n}}}\left(\bar{\xi}_{\vec{n}}P\right)
\\+\frac{1}{2}\sum_{\vec{n}}\frac{\partial}{\partial\bar{\xi}_{\vec{n}}}\left[P\sum_{\vec{\ell}_{1}\ne\vec{n}}\sum_{\vec{\ell}_{2}}\sum_{\vec{\ell}_{3}}V_{\vec{n},\vec{\ell}_{1},\vec{\ell}_{2},\vec{\ell}_{3}}\bar{\xi}_{\vec{\ell}_{1}}\bar{\xi}_{\vec{\ell}_{2}}\bar{\xi}_{\vec{\ell}_{3}}\right]\,,
\end{multline}
where we have denoted $\bar{\xi}_{\vec{n}}=\bar{\xi}_{\vec{n}}(\bar{t}\,)$ for the sake of conciseness. Multiplying this equation by any functional $\mathbb{F}\left(\left\{ \bar{\xi}_{\vec{n}}\right\} \right)$ of the Fourier components of the field $\xi$ and integrating over all $\bar{\xi}_{\vec{n}}$ gives after some integration by parts
\begin{multline}
\frac{\partial\left\langle \mathbb{F}\right\rangle }{\partial\bar{t}}=\frac{1}{2}\sum_{\vec{n}}\left|\vec{n}\right|^{2}\left\langle \frac{\partial^{2}\mathbb{F}}{\partial\bar{\xi}_{\vec{n}}\partial\bar{\xi}_{-\vec{n}}}\right\rangle -g\sum_{\vec{n}}\left|\vec{n}\right|^{4}\left\langle \frac{\partial\mathbb{F}}{\partial\bar{\xi}_{\vec{n}}}\bar{\xi}_{\vec{n}}\right\rangle
\\-\frac{1}{2}\sum_{\vec{n}}\sum_{\vec{\ell}_{1}\ne\vec{n}}\sum_{\vec{\ell}_{2}}\sum_{\vec{\ell}_{3}}V_{\vec{n},\vec{\ell}_{1},\vec{\ell}_{2},\vec{\ell}_{3}}\left\langle \frac{\partial\mathbb{F}}{\partial\bar{\xi}_{\vec{n}}}\bar{\xi}_{\vec{\ell}_{1}}\bar{\xi}_{\vec{\ell}_{2}}\bar{\xi}_{\vec{\ell}_{3}}\right\rangle
\label{eq:FP moments}
\end{multline}
where we have defined the expectation values
\begin{equation}
\left\langle \mathbb{F}\right\rangle =\int\prod_{\vec{n}}d\bar{\xi}_{\vec{n}}\,\mathbb{F}\left(\left\{ \bar{\xi}_{\vec{n}}\right\} \right)P\left(\left\{ \bar{\xi}_{\vec{n}}\right\} ,\bar{t}\,\right)\,.
\end{equation}
In this paper we will only consider static quantities and thus the left-hand side of equation (\ref{eq:FP moments}) will vanish. When the function $\mathbb{F}\left(\left\{ \bar{\xi}_{\vec{n}}\right\} \right)$ is chosen such that $\left\langle \mathbb{F}\right\rangle$ is a moment of $P\left(\left\{ \bar{\xi}_{\vec{n}}\right\} ,\bar{t}\,\right)$, equation (\ref{eq:FP moments}) becomes an equation relating different moments to each other. For example, for $\mathbb{F}=\bar{\xi}_{\vec{n}}\bar{\xi}_{\vec{n}'}$, equation (\ref{eq:FP moments}) reads
\begin{multline}
0=\left|\vec{n}\right|^{2}\delta_{\vec{n},-\vec{n}'}-g\left(\left|\vec{n}\right|^{4}+\left|\vec{n}'\right|^{4}\right)\left\langle \bar{\xi}_{\vec{n}}\bar{\xi}_{\vec{n}'}\right\rangle \\
-\frac{1}{2}\sum_{\vec{\ell}_{2}}\sum_{\vec{\ell}_{3}}\left[\sum_{\vec{\ell}_{1}\ne\vec{n}}V_{\vec{n},\vec{\ell}_{1},\vec{\ell}_{2},\vec{\ell}_{3}}\left\langle \bar{\xi}_{\vec{n}'}\bar{\xi}_{\vec{\ell}_{1}}\bar{\xi}_{\vec{\ell}_{2}}\bar{\xi}_{\vec{\ell}_{3}}\right\rangle \right.\\
\left.+\sum_{\vec{\ell}_{1}\ne\vec{n}'}V_{\vec{n}',\vec{\ell}_{1},\vec{\ell}_{2},\vec{\ell}_{3}}\left\langle \bar{\xi}_{\vec{n}}\bar{\xi}_{\vec{\ell}_{1}}\bar{\xi}_{\vec{\ell}_{2}}\bar{\xi}_{\vec{\ell}_{3}}\right\rangle \right]\,,
\end{multline}
which expresses the second moment $\left\langle \bar{\xi}_{\vec{n}}\bar{\xi}_{\vec{n}'}\right\rangle$  in terms of the fourth moment $\left\langle \bar{\xi}_{\vec{n}}\bar{\xi}_{\vec{\ell}_{1}}\bar{\xi}_{\vec{\ell}_{2}}\bar{\xi}_{\vec{\ell}_{3}}\right\rangle$. We could obtain an expression for the fourth moments by subbing $\mathbb{F}=\bar{\xi}_{\vec{n}_{1}}\bar{\xi}_{\vec{n}_{2}}\bar{\xi}_{\vec{n}_{3}}\bar{\xi}_{\vec{n}_{4}}$ into equation (\ref{eq:FP moments}) though the resulting expression would need knowledge of the sixth moment. Reminiscent of the BBGKY hierarchy \cite{Balescu1975Book, PlischkeBergersen1994Book, Kardar2007Book}, this property is generic and deciding how to implement closure to these equations is a complicated question. Ideally, we would like to argue that the higher order moments only contribute at a higher order such that they can be neglected at zeroth order and used to perturbatively correct subsequent orders. As discussed at the end of section~\ref{sec:derivation}, the strong coupling of the nonlinear term guarantees that such a perturbative expansion is destined to fail.

\subsection{The Self-Consistent Expansion}

Instead, the SCE methodology argues that though we cannot neglect the higher order moments relative to the lower order ones, there might exist some other linear theory which we can expand around instead. That is, we can define a linear operator 
\begin{equation}
\mathcal{O}_{0}P=\frac{1}{2}\sum_{\vec{n}}\left|\vec{n}\right|^{2}\frac{\partial^{2}P}{\partial\bar{\xi}_{\vec{n}}\partial\bar{\xi}_{-\vec{n}}}+\sum_{\vec{n}}\Gamma_{\left|\vec{n}\right|}\frac{\partial}{\partial\bar{\xi}_{\vec{n}}}\left(\bar{\xi}_{\vec{n}}P\right)
\end{equation}
where $\Gamma_{\left|\vec{n}\right|}$ is a free parameter which will be determined in the continuation and rewrite equation (\ref{eq:FP operator}) as
\begin{equation}
\frac{\partial P}{\partial\bar{t}}=\mathcal{O}_{0}P + \left(\mathcal{O}-\mathcal{O}_{0}\right)P\,.
\end{equation}
At this stage, one can think of $\Gamma_{\left|\vec{n}\right|}$ as an unknown renormalisation of the the bending rigidity, related to $\kappa_R(\mathbf{q}) $ in \cite{Nelson1987}, and its precise significance will become apparent as the calculation proceeds.

Explicitly in terms of moments, this equation becomes
\begin{multline}
0=\frac{1}{2}\sum_{\vec{n}}\left|\vec{n}\right|^{2}\left\langle \frac{\partial^{2}\mathbb{F}}{\partial\bar{\xi}_{\vec{n}}\partial\bar{\xi}_{-\vec{n}}}\right\rangle-\sum_{\vec{n}}\Gamma_{\left|\vec{n}\right|}\left\langle \frac{\partial\mathbb{F}}{\partial\bar{\xi}_{\vec{n}}}\bar{\xi}_{\vec{n}}\right\rangle\\ 
-\sum_{\vec{n}}\left(g\left|\vec{n}\right|^{4}-\Gamma_{\left|\vec{n}\right|}\right)\left\langle \frac{\partial\mathbb{F}}{\partial\bar{\xi}_{\vec{n}}}\bar{\xi}_{\vec{n}}\right\rangle \\
-\frac{1}{2}\sum_{\vec{n}}\sum_{\vec{\ell}_{1}\ne\vec{n}}\sum_{\vec{\ell}_{2}}\sum_{\vec{\ell}_{3}}V_{\vec{n},\vec{\ell}_{1},\vec{\ell}_{2},\vec{\ell}_{3}}\left\langle \frac{\partial\mathbb{F}}{\partial\bar{\xi}_{\vec{n}}}\bar{\xi}_{\vec{\ell}_{1}}\bar{\xi}_{\vec{\ell}_{2}}\bar{\xi}_{\vec{\ell}_{3}}\right\rangle\,.
\end{multline}
$\Gamma_{\left|\vec{n}\right|}$ can be thought of as an effective coupling constant such that corrections to this linear theory are kept small. The value of $\Gamma_{\left|\vec{n}\right|}$ will be determined in the continuation though due to the isotropic character of our system, we have however already assumed that it can only depend on the size of $\vec{n}$ and not its direction. Now if $\left\langle \mathbb{F}\right\rangle ^{\left(m\right)}$ denotes an $m^{th}$ order expansion of $\left\langle \mathbb{F}\right\rangle$, then by assumption, the latter terms which follow from the nonlinear operator $\left(\mathcal{O}-\mathcal{O}_0\right)$ will contribute at a higher order and thus we can devise an iterative scheme
\begin{multline}
0=\frac{1}{2}\sum_{\vec{n}}\left|\vec{n}\right|^{2}\left\langle \frac{\partial^{2}\mathbb{F}}{\partial\bar{\xi}_{\vec{n}}\partial\bar{\xi}_{-\vec{n}}}\right\rangle ^{\left(m\right)}-\sum_{\vec{n}}\Gamma_{\left|\vec{n}\right|}\left\langle \frac{\partial\mathbb{F}}{\partial\bar{\xi}_{\vec{n}}}\bar{\xi}_{\vec{n}}\right\rangle ^{\left(m\right)}\\
-\sum_{\vec{n}}\left(g\left|\vec{n}\right|^{4}-\Gamma_{\left|\vec{n}\right|}\right)\left\langle \frac{\partial\mathbb{F}}{\partial\bar{\xi}_{\vec{n}}}\bar{\xi}_{\vec{n}}\right\rangle ^{\left(m-1\right)}\\
-\frac{1}{2}\sum_{\vec{n}}\sum_{\vec{\ell}_{1}\ne\vec{n}}\sum_{\vec{\ell}_{2}}\sum_{\vec{\ell}_{3}}V_{\vec{n},\vec{\ell}_{1},\vec{\ell}_{2},\vec{\ell}_{3}}\left\langle \frac{\partial\mathbb{F}}{\partial\bar{\xi}_{\vec{n}}}\bar{\xi}_{\vec{\ell}_{1}}\bar{\xi}_{\vec{\ell}_{2}}\bar{\xi}_{\vec{\ell}_{3}}\right\rangle ^{\left(m-1\right)}
\label{eq:FP SCE}
\end{multline}
relating higher order moments to lower order ones.
This equation can now be used to obtain any moment up to any order. For example, to obtain the second moment at zeroth order, it is sufficient to sub $\mathbb{F}=\bar{\xi}_{\vec{n}}\bar{\xi}_{\vec{n}'}$ and $m=0$ into equation (\ref{eq:FP SCE}), in which case the $(m-1)$ terms drop out and one immediately obtains
\begin{equation}
\left\langle \bar{\xi}_{\vec{n}}\bar{\xi}_{\vec{n}'}\right\rangle ^{\left(0\right)}=\frac{\left|\vec{n}\right|^{2}}{2\Gamma_{\left|\vec{n}\right|}}\delta_{\vec{n},-\vec{n}'}\,.
\label{eq:2pt 0 order}
\end{equation}
Similarly, subbing in $\mathbb{F}=\bar{\xi}_{\vec{n}_{1}}\bar{\xi}_{\vec{n}_{2}}\bar{\xi}_{\vec{n}_{3}}\bar{\xi}_{\vec{n}_{4}}$ and $m=0$ ultimately gives the four-point function at zeroth order as
\begin{multline}
\left\langle \bar{\xi}_{\vec{n}_{1}}\bar{\xi}_{\vec{n}_{2}}\bar{\xi}_{\vec{n}_{3}}\bar{\xi}_{\vec{n}_{4}}\right\rangle ^{\left(0\right)}=\left\langle \bar{\xi}_{\vec{n}_{1}}\bar{\xi}_{\vec{n}_{2}}\right\rangle ^{\left(0\right)}\left\langle \bar{\xi}_{\vec{n}_{3}}\bar{\xi}_{\vec{n}_{4}}\right\rangle ^{\left(0\right)}\\
+\left\langle \bar{\xi}_{\vec{n}_{1}}\bar{\xi}_{\vec{n}_{3}}\right\rangle ^{\left(0\right)}\left\langle \bar{\xi}_{\vec{n}_{2}}\bar{\xi}_{\vec{n}_{4}}\right\rangle ^{\left(0\right)}+\left\langle \bar{\xi}_{\vec{n}_{1}}\bar{\xi}_{\vec{n}_{4}}\right\rangle ^{\left(0\right)}\left\langle \bar{\xi}_{\vec{n}_{2}}\bar{\xi}_{\vec{n}_{3}}\right\rangle ^{\left(0\right)}\,,
\end{multline}
which is just Isserlis' theorem \cite{Isserlis1916, Isserlis1918}, also known as Wick's theorem \cite{Kardar2007Book}, for the four-point function of a multivariate Gaussian distribution.

\begin{figure}
\includegraphics[width=0.9\columnwidth]{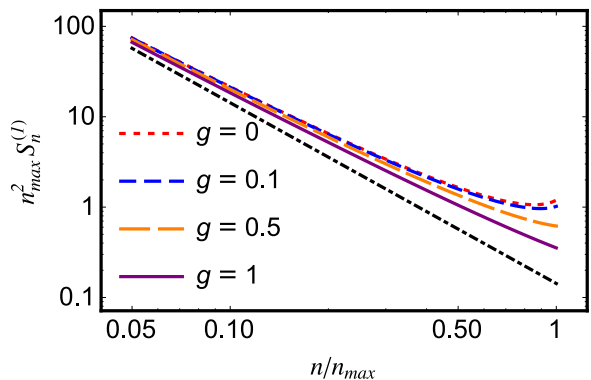}%
\caption{The first order approximation for the structure factor $S_{\vec{n}}$ of the sheet as given by equation (\ref{eq:S_n}) for various values of $g$ (solid and dashed lines). The dot-dashed line is a guideline for the power law $\sim n^{-2}$. For small $n$, the logarithmic correction can be seen to introduce a small deviation from the power law, while for large $n$, the polynomial correction $\left[1+B\left(g\right)\frac{n}{n_{max}}\right]$ introduces a small increasing tail which becomes more pronounced with shrinking $g$.
\label{fig:S_n theory}}
\end{figure}

To study the effect of the nonlinearity, we proceed to higher orders. Subbing in $\mathbb{F}=\bar{\xi}_{\vec{n}}\bar{\xi}_{\vec{n}'}$ and $m=1$ gives the following expression for the two-point function at first order
\begin{multline}
\left(\Gamma_{\left|\vec{n}\right|}+\Gamma_{\left|\vec{n}'\right|}\right)\left\langle \bar{\xi}_{\vec{n}}\bar{\xi}_{\vec{n}'}\right\rangle ^{\left(1\right)}=\left|\vec{n}\right|^{2}\delta_{\vec{n},-\vec{n}'}\\
-2\left(g\left|\vec{n}\right|^{4}-\Gamma_{\left|\vec{n}\right|}\right)\left\langle \bar{\xi}_{\vec{n}}\bar{\xi}_{\vec{n}'}\right\rangle ^{\left(0\right)}\\
-\frac{1}{2}\sum_{\begin{subarray}{c}
\vec{\ell}_{1}\ne\vec{n}\\
\vec{\ell}_{2},\vec{\ell}_{3}
\end{subarray}}V_{\vec{n},\vec{\ell}_{1},\vec{\ell}_{2},\vec{\ell}_{3}}\left\langle \bar{\xi}_{\vec{n}'}\bar{\xi}_{\vec{\ell}_{1}}\bar{\xi}_{\vec{\ell}_{2}}\bar{\xi}_{\vec{\ell}_{3}}\right\rangle ^{\left(0\right)}\\
-\frac{1}{2}\sum_{\begin{subarray}{c}
\vec{\ell}_{1}\ne\vec{n}'\\
\vec{\ell}_{2},\vec{\ell}_{3}
\end{subarray}}V_{\vec{n}',\vec{\ell}_{1},\vec{\ell}_{2},\vec{\ell}_{3}}\left\langle \bar{\xi}_{\vec{n}}\bar{\xi}_{\vec{\ell}_{1}}\bar{\xi}_{\vec{\ell}_{2}}\bar{\xi}_{\vec{\ell}_{3}}\right\rangle ^{\left(0\right)}
\end{multline}
where we have already found expressions for the zeroth order two-point and four-point functions. Subbing in these expressions together with the kernel $V_{\vec{n},\vec{\ell}_{1},\vec{\ell}_{2},\vec{\ell}_{3}}$ allows us to simplify the sums to obtain
\begin{multline}
\left\langle \bar{\xi}_{\vec{n}}\bar{\xi}_{\vec{n}'}\right\rangle ^{\left(1\right)}=\left\langle \bar{\xi}_{\vec{n}}\bar{\xi}_{\vec{n}'}\right\rangle ^{\left(0\right)}+\frac{\left\langle \bar{\xi}_{\vec{n}}\bar{\xi}_{\vec{n}'}\right\rangle ^{\left(0\right)}}{\Gamma_{\left|\vec{n}\right|}}\times\\
\times\left(\Gamma_{\left|\vec{n}\right|}-g\left|\vec{n}\right|^{4}-\frac{1}{2}\sum_{\vec{\ell}\ne\vec{n}}\frac{\left|\vec{\ell}\,\right|^{2}\left|\vec{n}\times\vec{\ell}\,\right|^{4}}{\Gamma_{\left|\vec{\ell}\,\right|}\left|\vec{n}-\vec{\ell}\,\right|^{4}}\right)\,.
\end{multline}
This equation expresses the two-point function at first order as a correction to the zeroth order two-point function. If we wish our expansion for the two-point function to be meaningful, it must be the case that corrections to the zeroth order approximation be small. Accordingly, self-consistency demands that we select $\Gamma_{\left|\vec{n}\right|}$ so as to ensure that this be the case however since we have complete freedom to choose $\Gamma_{\left|\vec{n}\right|}$, we can go even further and demand that the first order correction vanish entirely. There are of course other ways to select $\Gamma_{\left|\vec{n}\right|}$ however previous work \cite{Schwartz2008} has shown this method to be extremely successful as it introduces closure into our hierarchy of moment equations in a manner  that ensures they are self-consistently correct up to first order. Though we will not need to do so here, this approach can be applied at any order to obtain a corresponding self-consistent $m^{th}$ order theory. Each such theory will give a different expression for $\Gamma_{\left|\vec{n}\right|}$ thus $\Gamma_{\left|\vec{n}\right|}$ is not strictly an effective coupling constant or renormalised bending rigidity but rather the coefficient of a particular linear theory whose $m^{th}$ order expansion closely resembles the $m^{th}$ order expansion of the nonlinear theory we are interested in. For the first order theory we are developing here, we obtain
\begin{equation}
\Gamma_{\left|\vec{n}\right|}=g\left|\vec{n}\right|^{4}+\frac{1}{2}\sum_{\vec{\ell}\ne\vec{n}}\frac{\left|\vec{\ell}\,\right|^{2}\left|\vec{n}\times\vec{\ell}\,\right|^{4}}{\Gamma_{\left|\vec{\ell}\,\right|}\left|\vec{n}-\vec{\ell}\,\right|^{4}}\,,
\label{eq:disc integ equ}
\end{equation}
which is a two-dimensional nonlinear discrete integral equation for $\Gamma_{\left|\vec{n}\right|}$. While solving such equations is in general extremely challenging, in this particular case, the equation is indeed amenable to analytic methods. In the appendix, we provide a detailed derivation of the approximate solution
\begin{multline}
\Gamma_{\left|\vec{n}\right|}\simeq n^{4}\sqrt{\frac{3\pi}{4}\left[A\left(g\right)-\ln\left(\frac{n}{n_{max}}\right)\right]}\times\\
\times\left[1+B\left(g\right)\frac{n}{n_{max}}\right]\,,
\end{multline}
where $n_{max}$ is some upper-frequency cut-off which must be imposed on the equation. $A\left(g\right)$ and $B\left(g\right)$ are constants which only depend on $g$ and are determined by the implicit solutions to equations~(\ref{eq:A,B 1}) and (\ref{eq:A,B 2}) in the appendix. Since the physically interesting case is $g\ll1$, it is more convenient to provide the small $g$ expansions
\begin{eqnarray}
A&\simeq &0.137+0.336g+0.243g^{2}+0.112g^3+O\left(g^{4}\right), \label{eq:A theory} \\
B&\simeq &-0.265+0.360g-0.395g^{2}+0.311g^3+O\left(g^{4}\right).\qquad \label{eq:B theory}
\end{eqnarray}
Combining with equation (\ref{eq:2pt 0 order}), we obtain that the structure factor up to first order is
\begin{multline}
\left\langle \bar{\xi}_{\vec{n}}\bar{\xi}_{\vec{n}'}\right\rangle ^{\left(1\right)}=\\
\frac{\delta_{\vec{n},-\vec{n}'}}{n^{2}\sqrt{3\pi\left[A\left(g\right)-\ln\left(\frac{n}{n_{max}}\right)\right]}\left[1+B\left(g\right)\frac{n}{n_{max}}\right]}\,.
\label{eq:S_n}
\end{multline}
In Fig.~\ref{fig:S_n theory}, this structure factor $S_n^{(1)} = \left\langle \bar{\xi}_{\vec{n}}\bar{\xi}_{-\vec{n}}\right\rangle ^{\left(1\right)}$ is plotted for various values of $g$ together with a guide line for $\sim n^{-2}$. Since the log-log plots appear straight over large intervals, they give the illusion of a power law with anomalous exponent close to but not exactly $-2$. In actual fact, it is the logarithmic and polynomial corrections which causes the deviation from the non-anomalous power law. While for small $n$, all values of $g$ tend towards the same asymptote, for large $n$, the polynomial correction, determined by the coefficient $B\left(g\right)$, becomes increasingly important, even giving rise to a small increasing tail for extremely small values of $g$.

Before moving to the next section, we would like to discuss the renormalisation of the coupling constant $g$.  Simple power counting of the linear terms shows that the nonlinear quartic interaction scales as $L^{2-d}$, where $d$ is the dimensionality of the sheet. Since $d=2$ for the problem of interest, the nonlinear term is marginal and its relevance or irrelevance at long distances can only be decided by a renormalization group calculation. This point is also discussed in \cite{Nelson1987} where the assumption that the coupling constant is not renormalised is assumed. It turns out that the self-consistent expansion does not indicate any renormalisation of the coupling constant $g$ itself -- at least not at the leading order used in this paper. It is however important to stress that this option cannot be ruled out.

\section{Comparison with Simulations\label{sec:simulations}}

\begin{figure}
\includegraphics[width=0.8\columnwidth]{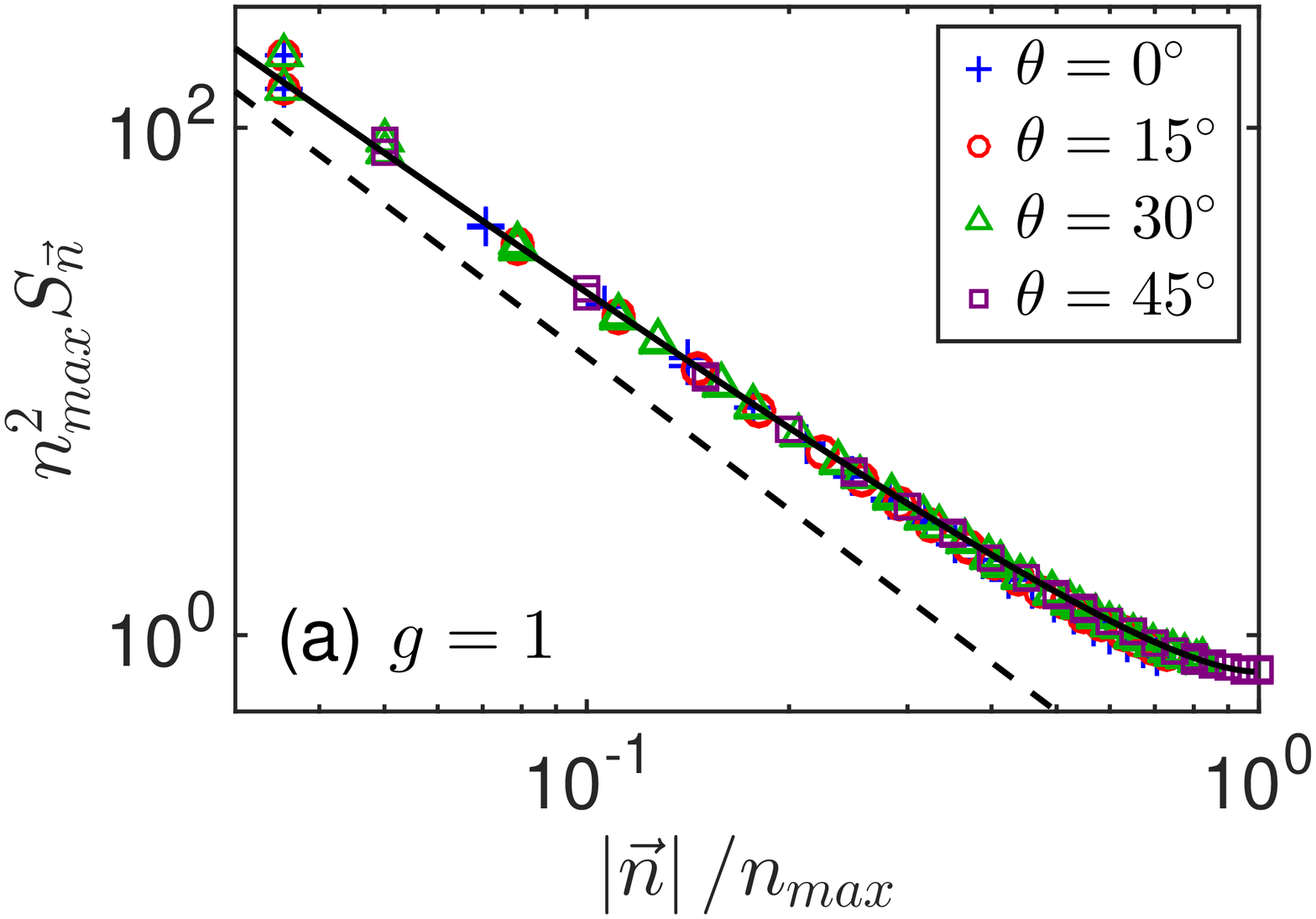}\\%
\includegraphics[width=0.8\columnwidth]{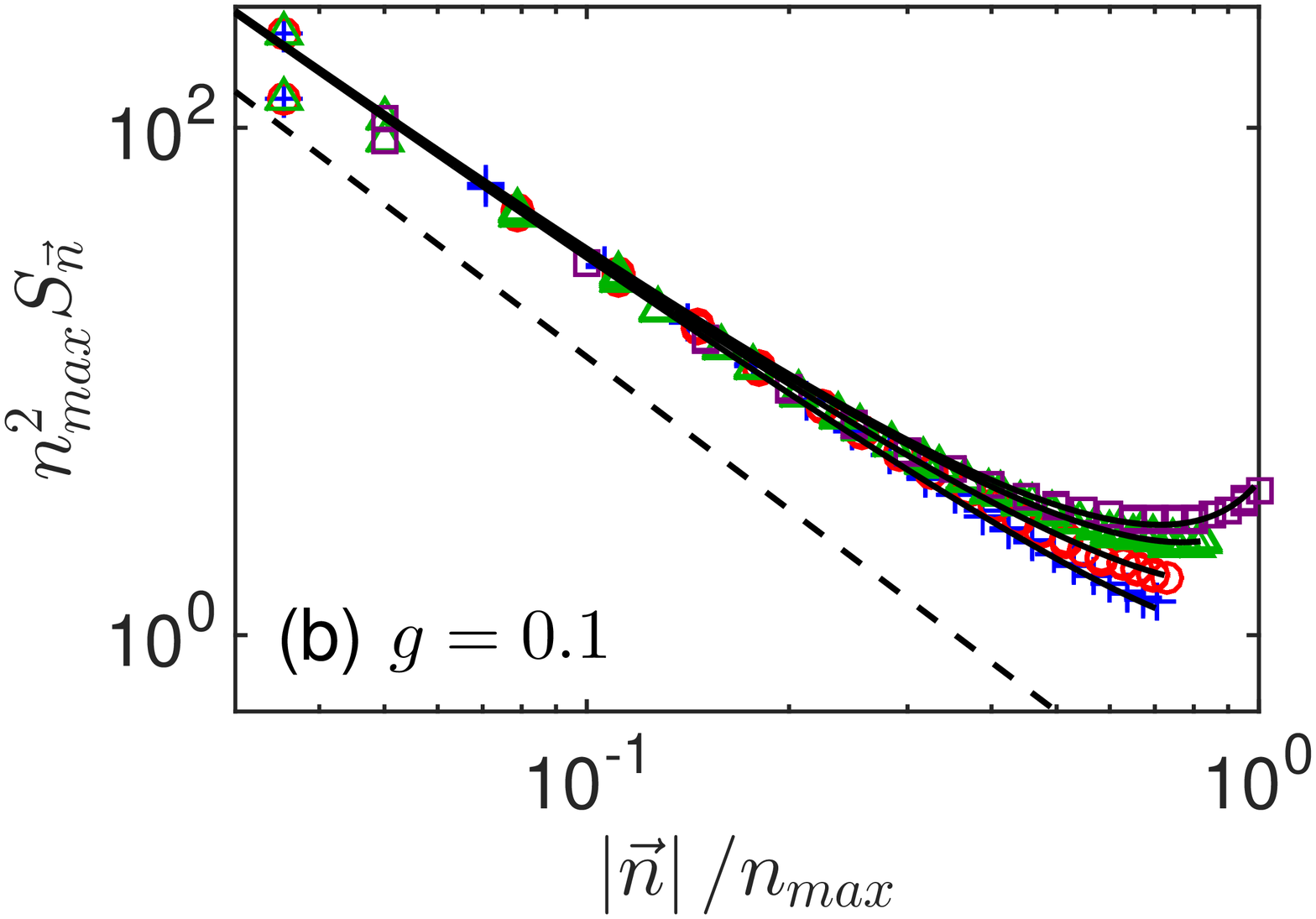}\\%
\includegraphics[width=0.8\columnwidth]{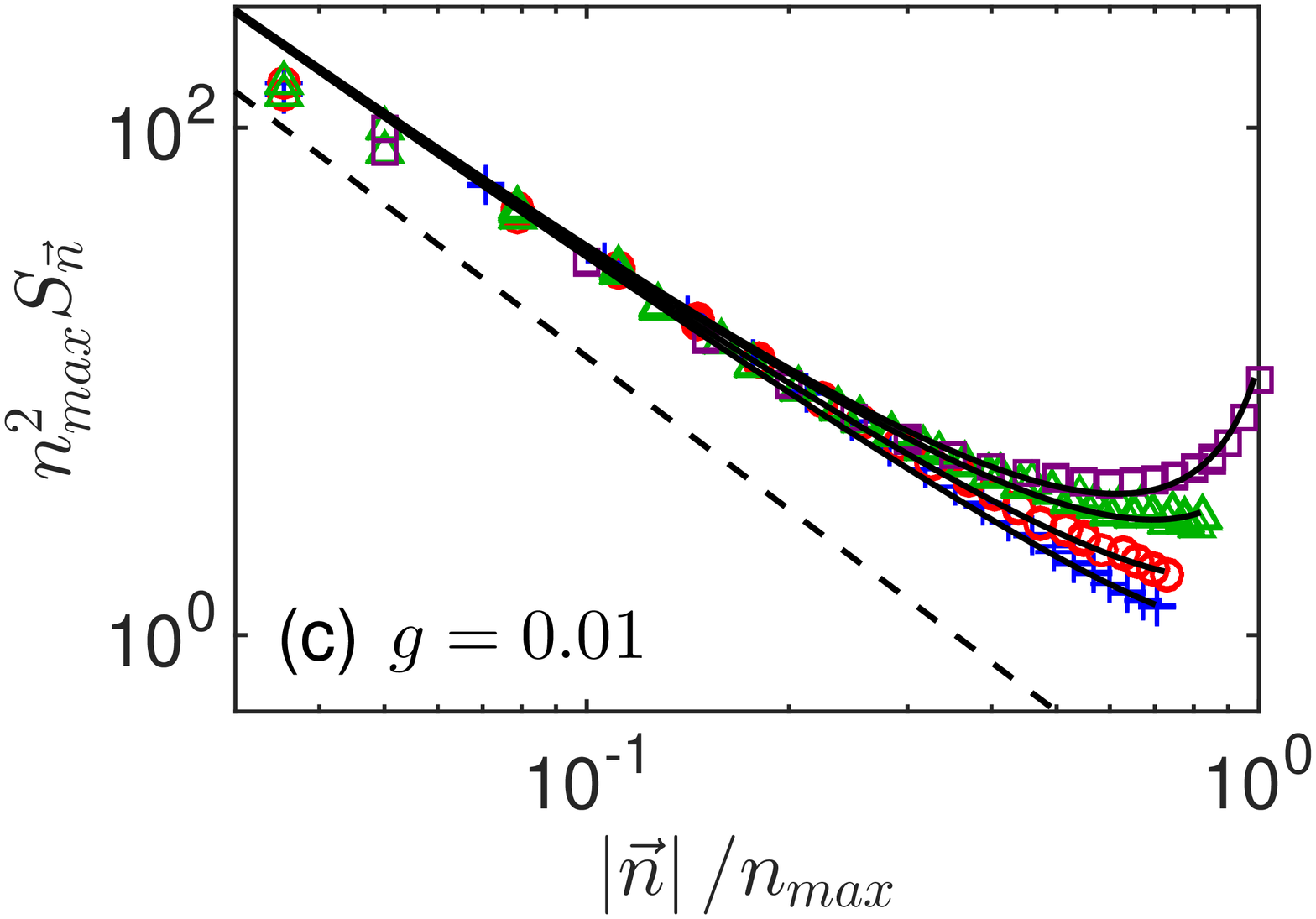}\\%
\includegraphics[width=0.8\columnwidth]{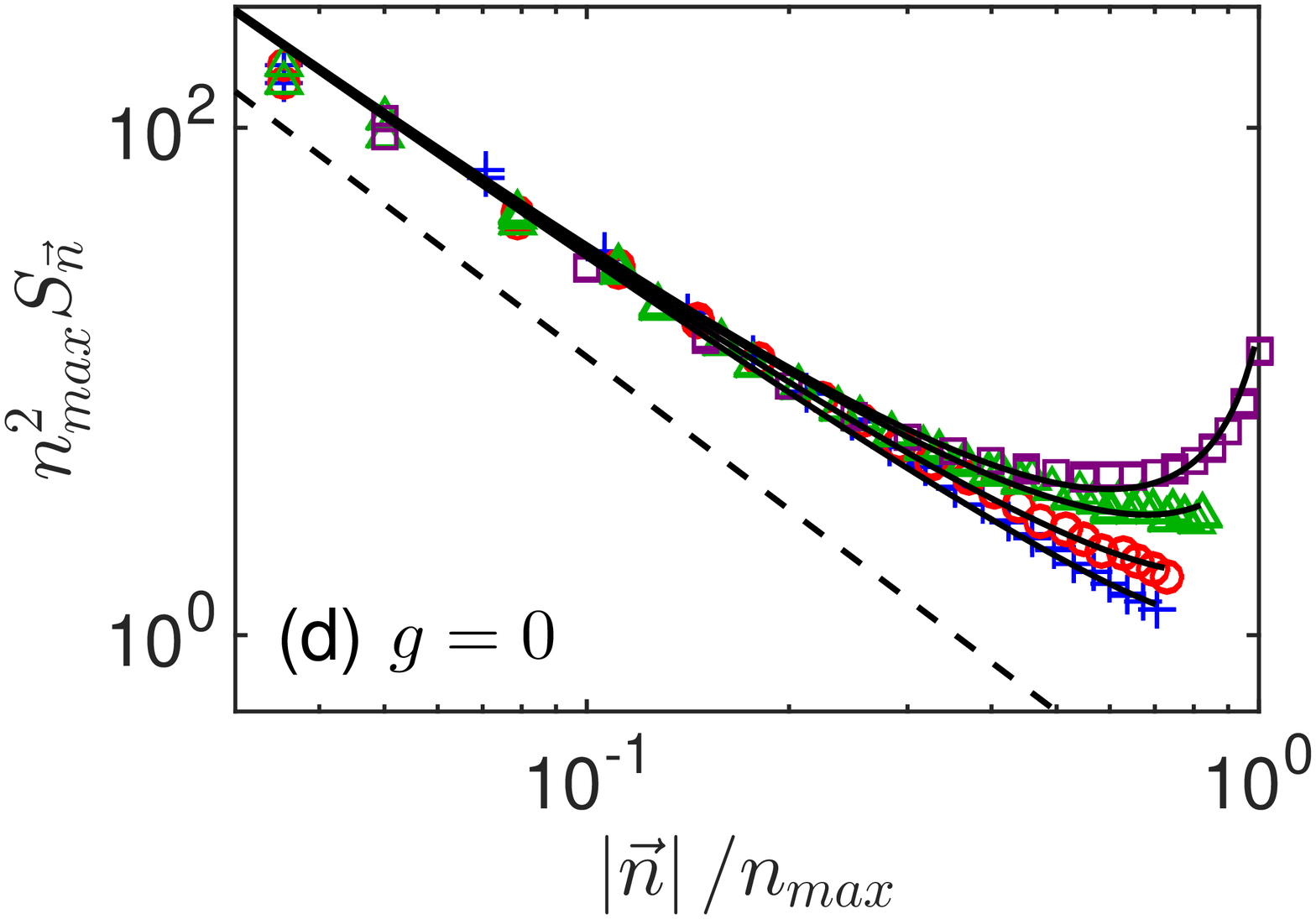}%
\caption{Numerical results for the structure factor $S_{\vec{n}}=\left\langle \bar{\xi}_{\vec{n}}\bar{\xi}_{-\vec{n}}\right\rangle$ derived from simulating equation (\ref{eq:fvk fourier dmnless}) with various values of $g$: (a) $g=1$, (b) $g=0.1$, (c) $g=0.01$, (d) $g=0$. The simulations are not completely isotropic thus for each simulation, the structure factor has been computed at angles of $0^{\circ}$ (blue plus signs), $15^{\circ}$ (red circles), $30^{\circ}$ (green triangles) and $45^{\circ}$ (purple squares) relative to the ``x-axis''. The solid black lines are fits of equation (\ref{eq:S_n}) to each data set while the dashed lines are guidelines proportional to $n^{-2}$.
\label{fig:S_n Experimental}}
\end{figure}

Equation (\ref{eq:fvk fourier dmnless}), which describes the fluctuations of the sheet height in Fourier space, is an ordinary Langevin field equation and thus in principle can be simulated over say a square lattice by forward iteration with a small discretisation of the time step. In practice, the conserved noise can easily be computed in the Fourier space but computing the quartic interaction term directly can be time consuming. To circumvent this problem, a pseudo-spectral method, as described in \cite{During2017}, can be implemented. This entails expressing the quartic interaction as the Fourier transform of its real space counterpart which can be calculated far more quickly though this also has the effect of imposing periodic boundary conditions on our fluctuating sheet. The fact that a maximum frequency $n_{max}$ must be chosen to perform simulations poses no problem as our solution, given by equation (\ref{eq:S_n}), indeed assumes the existence of such a cut-off. Physically, this corresponds to the fact that no sheet can be probed with infinite resolution. The time step $\delta t$ however must be taken to be sufficiently small to avoid numerical instability. The maximum size $\delta t$ can be taken while ensuring stability is determined by the maximum frequency used, with larger values of $n_{max}$ requiring smaller time steps. Accordingly $n_{max}$ cannot be too large if we wish to simulate equation (\ref{eq:fvk fourier dmnless}) over sufficiently long periods of time to obtain meaningful results for the structure factor $S_{\vec{n}}$. The results shown in Fig.~\ref{fig:S_n Experimental} were obtained for the frequencies $\vec{n}\in\left[-20,20\right]\times\left[-20,20\right]$, ie. a $41\times41$ square lattice such that the maximum frequency is $n_{max}=20\sqrt{2}\approx28$. To obtain numerically stable results, a time step of $\delta t=10^{-6}$ was used and for each value of $g\in\left\{ 1,0.1,0.01,0\right\}$, $10$ simulations were run over $10^{6}$ time steps. 

To validate the functional form of the structure factor $S_{\vec{n}}$ predicted by our theory, equation (\ref{eq:S_n}) with $A\left(g\right)$ and $B\left(g\right)$ as free parameters was fitted (solid black lines) to each data set shown in Fig.~\ref{fig:S_n Experimental} (coloured shapes). It can be seen that no matter the value of $g$, equation (\ref{eq:S_n}) indeed captures the shape of the structure factor. The dashed line in each figure is a guideline proportional to $n^{-2}$ from which it can easily be seen how the logarithmic correction in equation (\ref{eq:S_n}) slightly modifies the slope of the structure factor.

The periodic boundary conditions over a square lattice have the effect of breaking the isotropy of the system and thus the structure factor $S_{\vec{n}}$ is found to also depend on the angle of $\vec{n}$ and not just its magnitude. For this reason, in each figure, separate data has been plotted for angles of $0^{\circ}$, $15^{\circ}$, $30^{\circ}$ and $45^{\circ}$. Curiously, the degree of anisotropy decreases with increasing $g$ vanishing completely as $g$ approaches 1. Even though our theory is derived under the assumption of complete isotropy, the functional form of $S_{\vec{n}}$ remains a superb fit for each data set in the presence of the anisotropy. Indeed, for any given $g$, the anisotropy only seems to effect the tail parameter, $B\left(g\right)$, while the parameter $A\left(g\right)$, related to the logarithmic correction, is entirely unaffected by the angle of $\vec{n}$. This robustness to slightly anisotropic circumstances can be viewed as further validation of the functional form given by equation (\ref{eq:S_n}). 

\begin{figure}
\includegraphics[width=\columnwidth]{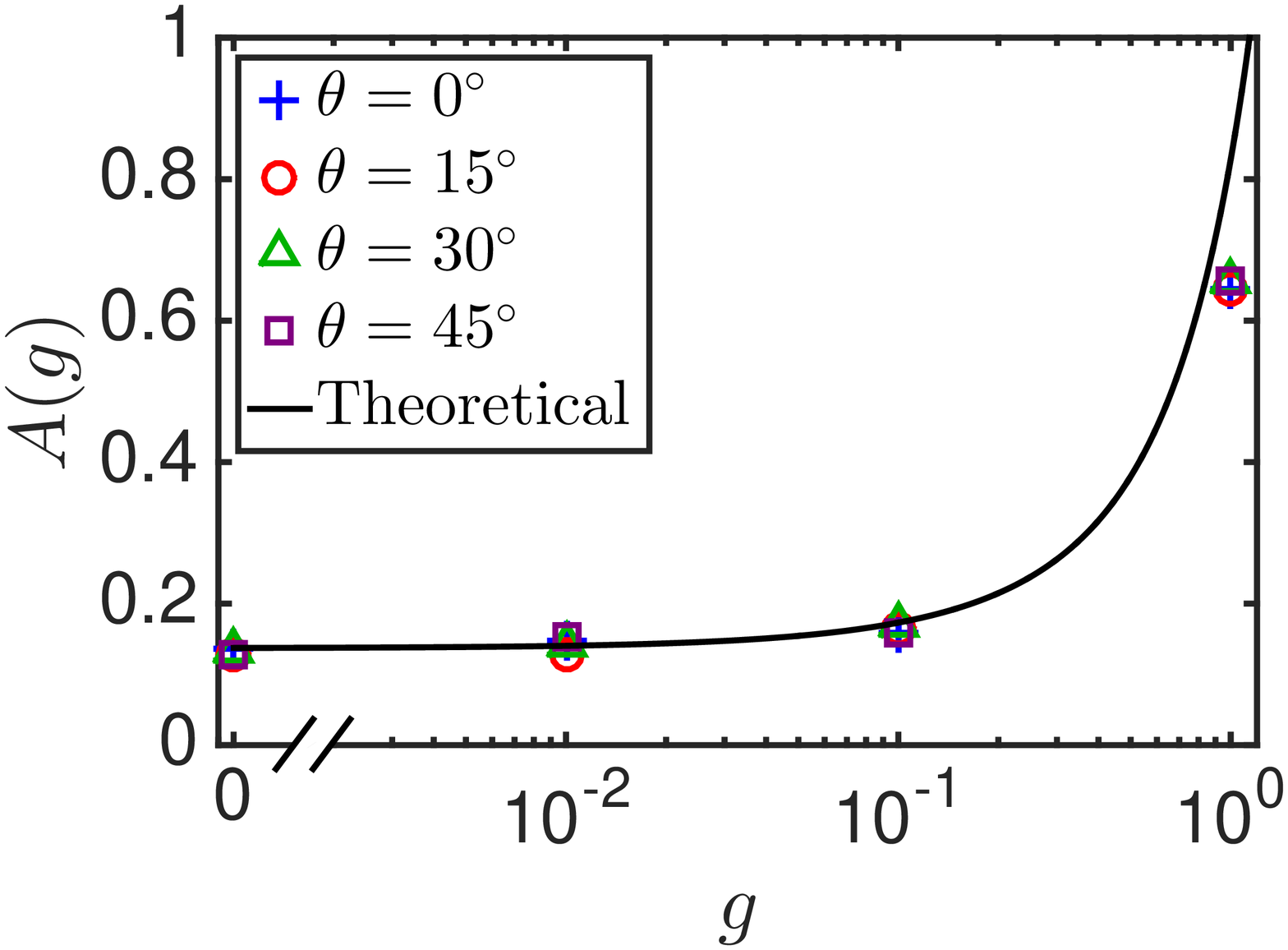}\\%
\includegraphics[width=\columnwidth]{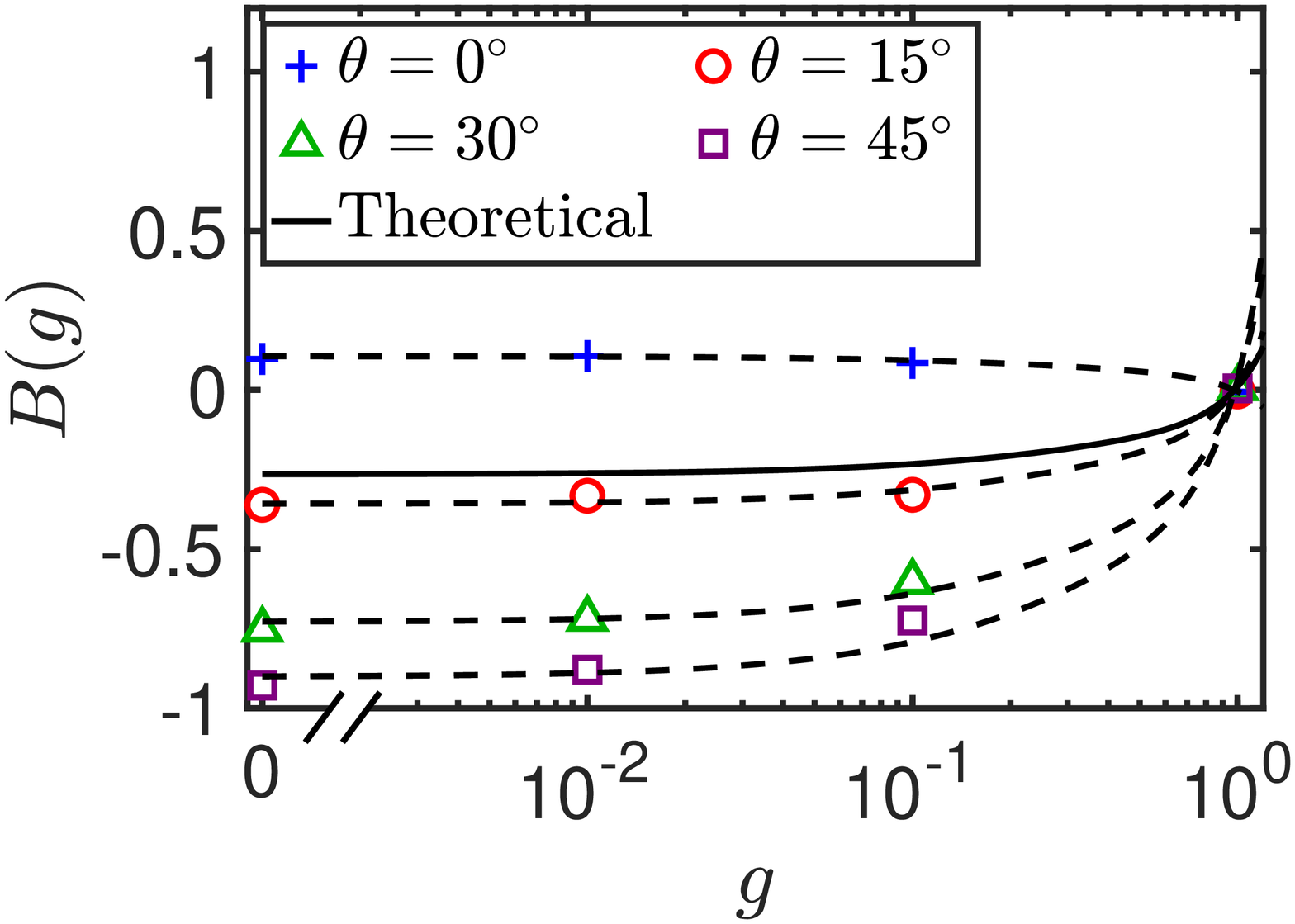}\\%
\caption{Comparisons between the theoretical values of $A(g)$ and $B(g)$ given by equations (\ref{eq:A theory}) and (\ref{eq:B theory}) (solid lines) and the values obtained from fitting equation (\ref{eq:S_n}) to the simulation data (various shapes). The dashed lines in the latter plot are angular dependent scalings $f(\theta)B(g)$ of equation (\ref{eq:B theory}). Equation (\ref{eq:A theory}) and the scaled forms of equation (\ref{eq:B theory}) beautifully predict the fitted value of $A(g)$ and $B(g)$ over a wide range of $g$ values.
\label{fig:A B comparisons}}
\end{figure}

Comparisons between the theoretical values of $A(g)$ and $B(g)$, given by equations (\ref{eq:A theory}) and (\ref{eq:B theory}), and the values obtained from fitting equation (\ref{eq:S_n}) to our simulation data are shown in Fig.~\ref{fig:A B comparisons}. The theoretical values predicted by equation (\ref{eq:A theory}) for $A(g)$ (solid line) can be seen to be extremely accurate over a wide range of $g$ values, only breaking down as $g$ approaches 1. In contrast, due to the anisotropic nature of $B(g)$, the theoretical values predicted by equation (\ref{eq:B theory}) (solid line) do not capture any of the particular fit values. One can however obtain superb fits by simply scaling the function $B(g)$ given by equation (\ref{eq:B theory}) by a constant, that is, by finding for each angle $\theta$ the quantity $f(\theta)$ which minimises the difference between $f(\theta)B(g)$ and the numerical results for $B(g)$ \footnote{We find $f(0^\circ) = -0.40$, $f(15^\circ) = 1.35$, $f(30^\circ) = 2.75$, $f(45^\circ) = 3.40$}. The dashed lines in Fig.~\ref{fig:A B comparisons} show these optimal scalings and can be seen to be excellent fits for all angles over the entire range of $g$ values. As mentioned above, the origin of this angular dependent scaling is due to the periodic boundary conditions used in our simulation though it is curious that this anisotropy only effects $B(g)$ and not $A(g)$. Though the particular nature of $f(\theta)$ is not fully understood, these results inform us that, at least in this case, a more precise expression for the structure factor would be
\begin{multline}
\left\langle \bar{\xi}_{\vec{n}}\bar{\xi}_{\vec{n}'}\right\rangle ^{\left(1\right)}=\\
\frac{\delta_{\vec{n},-\vec{n}'}}{n^{2}\sqrt{3\pi\left[A\left(g\right)-\ln\left(\frac{n}{n_{max}}\right)\right]}\left[1+B\left(g\right)\frac{n}{n_{max}/f(\theta)}\right]}
\end{multline}
where $A(g)$ and $B(g)$ are still given by equations (\ref{eq:A theory}) and (\ref{eq:B theory}) respectively. This expression suggests the angular scaling is essentially just a renormalisation of the maximum frequency supported by the sheet in each direction as a result of the boundary conditions. As elaborated on in the appendix, the function $B(g)$ is essentially determined by matching the general solution of equation (\ref{eq:disc integ equ}) with its large $\vec{n}$ boundary condition and thus it is perhaps unsurprising that an anisotropic renomalisation of the maximum supported frequency should imply a rescaling of $B(g)\frac{n}{n_{max}}$.

\section{Discussion\label{sec:discussion}}

We have combined the F\"oppl-von K\'arm\'an equations from elasticity theory with out-of-equilibrium statistical physics to develop a mathematical model for fluctuating thin sheets. Solving this model, we have shown that the structure factor of such sheets will follow nontrivial scaling -- a logarithmically corrected power law for small frequencies which becomes a rational function for large frequencies. The solution detailed in the appendix suggests that the problem requires an upper frequency cutoff and also provides definite relationships between the coefficients appearing in our solution and the physical properties such as sheet thickness $h$ and Young's modulus $E$. This opens up the possibility for experimental study through the measurement of such coefficients. 

Indeed, while experimental investigation of fluctuating sheets has been performed in the context of studying wave turbulence \cite{Boudaoud2008, Mordant2008, Cadot2008, Cobelli2009, Humbert2013}, we are not aware of any attempt to measure the structure factor of the fluctuating sheet itself. On the other hand, the structure factor of a crumpled sheet has been measured several times \cite{PlouraboueRoux1996, BlairKudrolli2005, Balankin2006, Andresen2007} and in each of these instances, anomalous power-laws have been loosely fitted to the data. Observed exponents cannot be simply explained by the structure factor of a single ridge $\sim 1/q^4$ and as random crumpling is merely the plastification of a particular set of random deformations of a thin sheet, it is not unreasonable to hypothesise a connection between the structure of a fluctuating sheet and that of a crumpled sheet. Accordingly, our results suggest that the structure factor of crumpled sheets might be better described by rational functions with logarithmic corrections, than by crude anomalous power-laws. The wide range of applications of fluctuating sheets including extremely thin sheets such as graphene suggests that many potential experiments to test our results can be developed. 

Furthermore, it is not only our results which have indicated the importance of logarithmic corrections to the structure factor and renormalised bending rigidity of thin sheets. In recent years, work carried out on thermally fluctuating sheets in the presence of external tension \cite{Kosmrlj2016, Shankar2021} have also observed logarithmic effects, though these are not identical with ours. In particular, our logarithmic correction occurs with a square root while theirs occurs unadorned. This prevalence amongst related yet distinct models is supportive of our argument that more sophisticated analyses which go beyond simple power laws are needed for a proper understanding of such systems.

Perhaps the biggest limitation of our theory is the absence of self-avoidance. As has been shown in \cite{Vliegenthart2006}, self-avoidance plays a significant role in the context of strong crumpling. On the other hand, in the context of weak crumpling, instances of self-avoidance will occur infrequently. Accordingly, neglecting self-avoidance simply limits the scope of our theory rather than its validity. 

As described in the introduction, much research into crumpled paper has focused on the formation of singular structures such as ridges and d-cones and their interactions. Since these structures are known to solve the F\"oppl-von K\'arm\'an equations \cite{Lobkovsky1997, BenAmar1997}, it is interesting to ask whether the approach taken in this paper can be connected directly to this research. For example, we wonder whether our approach can be reframed as some sort of many body model whose constituent parts are the various ridges and d-cones? If this were the case, our approach could be conceived of as a type of ``many crease'' generalisation of the theory of singular structures in thin sheets.

Further, we re-emphasise that the structure factor of fluctuating thin sheets is an input into much contemporary research including the mechanical and elastic properties of deformed sheets \cite{Kosmrlj2013, Kosmrlj2014, Kosmrlj2016, Kosmrlj2017, Shankar2021, Morshedifard2021} and their acoustic \cite{Kramer1996, Houle1996} and optical emissions \cite{Rad2019}. While such research has so far tended to assume ordinary power-laws, our results suggest that a wider family of inputs needs to be considered including rational functions, possibly with logarithmic corrections. Indeed, exploration of the effect of this wider range of functions on acoustic emissions could suggest novel questions of inference such as, to paraphrase Kac's provocative question \cite{Kac1966}, can one hear the shape of a fluctuating or crumpled sheet?

The approach we have taken in this paper naturally generalises to investigate other types of thin sheet and forcing. For example, instead of beginning with the flat sheet F\"oppl-von K\'arm\'an equations, we could begin with the F\"oppl-von K\'arm\'an equations for shells \cite{Paulose2012, Kosmrlj2017} or other types of curved surfaces \cite{VanHemmen2005} and this would allow us to investigate the effect of the underlying undeformed geometry on the fluctuating structure. Similarly, it would be of great interest to incorporate more realistic modelling of the forcing such as driving from the boundaries which would amount to complex spatio-temporal noise. Accordingly, the approach we have introduced here is extremely generic, providing a first step to analytically describing the behaviour of a wide range of different types of fluctuating surface.

\begin{acknowledgments}
The authors wish to thank one of the anonymous referees for insightful comments and analysis that helped to clarify the origin of the $\sqrt{\ln{q}}$ correction to the structure factor. This work was supported by the Israel Science Foundation Grant No. 1682/18.
\end{acknowledgments}

\appendix*
\section{Solution to the Integral Equation\label{sec:appendix}}

In this appendix, we outline a method for solving equation (\ref{eq:disc integ equ})
\begin{equation}
\Gamma_{\left|\vec{n}\right|}=g\left|\vec{n}\right|^{4}+\frac{1}{2}\sum_{\vec{\ell}\ne\vec{n}}\frac{\left|\vec{\ell}\,\right|^{2}\left|\vec{n}\times\vec{\ell}\,\right|^{4}}{\Gamma_{\left|\vec{\ell}\,\right|}\left|\vec{n}-\vec{\ell}\,\right|^{4}}\,,
\end{equation}
a two-dimensional nonlinear discrete integral equation for the parameter $\Gamma_{\left|\vec{n}\right|}$. To obtain an approximate solution, we can consider its continuous limit. In particular, let $\vec{q}=\vec{n}/L, \vec{k}=\vec{\ell}/L$ and $\Gamma\left(q\right)=\Gamma_{\left|\vec{n}\right|}/\left(\Lambda L\right)^{4}$ such that
\begin{equation}
\Gamma\left(q\right)=g\frac{\left|\vec{q}\,\right|^{4}}{\Lambda^{4}}+\frac{1}{2}\frac{1}{\Lambda^{8}}\sum_{\vec{k}\ne\vec{q}}\frac{1}{L^{2}}\frac{\left|\vec{k}\,\right|^{2}\left|\vec{q}\times\vec{k}\,\right|^{4}}{\Gamma\left(k\right)\left|\vec{q}-\vec{k}\,\right|^{4}}\,.
\end{equation}
Here, $\Lambda$ denotes some frequency scale which ensures that $\Gamma\left(q\right)$ remains dimensionless. In the large sheet limit of $L\rightarrow\infty$, the sum becomes an integral and we obtain a two-dimensional continuous integral equation for $\Gamma\left(q\right)$
\begin{equation}
\Gamma\left(q\right)=g\frac{\left|\vec{q}\,\right|^{4}}{\Lambda^{4}}+\frac{1}{2}\frac{1}{\Lambda^{8}}\int d^{2}k\frac{\left|\vec{k}\,\right|^{2}\left|\vec{q}\times\vec{k}\,\right|^{4}}{\Gamma\left(k\right)\left|\vec{q}-\vec{k}\,\right|^{4}}\,.
\end{equation}
Since $\Gamma\left(k\right)$ only depends on the size of $\vec{k}$ and not its direction, the angular part of the integral can be carried out exactly giving
\begin{equation}
\Gamma\left(q\right)=g\frac{q^{4}}{\Lambda^{4}}+\frac{3\pi}{8}\frac{1}{\Lambda^{8}}\left(\int_{0}^{q}dk\,\frac{k^{7}}{\Gamma\left(k\right)}+q^{4}\int_{q}^{\Lambda}dk\,\frac{k^{3}}{\Gamma\left(k\right)}\right)\,.
\label{eq:cont integ equ}
\end{equation}
Our task has thus been reduced to finding the solution to this one-dimensional continuous integral equation. In the continuation, it will become apparent that this integral equation only has a solution over finite intervals $q\in\left[0,q_{max}\right]$ where $q_{max}<\infty$ is some finite upper frequency cut-off. As such, we have chosen to identify the previously arbitrary frequency scale $\Lambda$ with the upper cut-off and introduced it into the integral bounds.

To solve this equation, differentiate it twice with respect to $q$ to obtain
\begin{equation}
\frac{d\Gamma}{dq}=4g\frac{q^{3}}{\Lambda^{4}}+\frac{3\pi}{2}\frac{q^{3}}{\Lambda^{8}}\int_{q}^{\Lambda}dk\,\frac{k^{3}}{\Gamma\left(k\right)}
\label{eq:cont integ equ dif}
\end{equation}
and
\begin{equation}
\frac{d^{2}\Gamma}{dq^{2}}=12g\frac{q^{2}}{\Lambda^{4}}+\frac{3\pi}{2}\frac{1}{\Lambda^{8}}\left(3q^{2}\int_{q}^{\Lambda}dk\,\frac{k^{3}}{\Gamma\left(k\right)}-\frac{q^{6}}{\Gamma\left(q\right)}\right)\,.
\end{equation}
These equations can now be combined to eliminate the integral giving rise to a second order nonlinear ODE
\begin{equation}
\frac{d^{2}\Gamma}{dq^{2}}=\frac{3}{q}\frac{d\Gamma}{dq}-\frac{3\pi}{2}\frac{1}{\Lambda^{8}}\frac{q^{6}}{\Gamma\left(q\right)}\,.
\label{eq:ODE}
\end{equation}
Apart from being far easier to solve than the integral equation, this equation has the added benefit that the previously explicit $g$ dependence has dropped out. Instead, the $g$ dependence now enters in the form of boundary conditions. In particular, the limits $q\rightarrow0$ in equation (\ref{eq:cont integ equ}) and $q\rightarrow\Lambda$ in equation (\ref{eq:cont integ equ dif}) give rise to the boundary conditions
\begin{eqnarray}
\Gamma\left(q=0\right)&=0\,,\label{eq:small q BC}\\
\left.\frac{d\Gamma}{dq}\right|_{q=\Lambda}&= \dfrac{4g}{\Lambda} \,.\label{eq:large q BC}
\end{eqnarray}
The general solutions to equation (\ref{eq:ODE}) are easily obtained numerically with arbitrary starting values. These solutions all exhibit movable singularities \cite{BenderOrszag2013Book} such that for sufficiently large $q$, $\Gamma\left(q\right)$ plummets to 0 and beyond this movable singularity, the solution is no longer defined. On this basis, the aforementioned upper frequency cut-off $\Lambda$ was identified.

\begin{figure}[t]
\includegraphics[width=0.8\columnwidth]{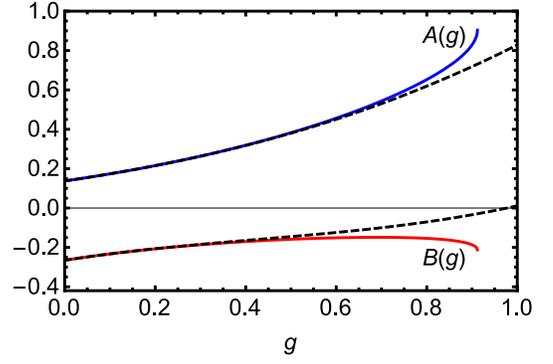}%
\caption{$A$ and $B$ as functions of $g$. The solid lines are the exact solution obtained from equations (\ref{eq:A,B 1}) and (\ref{eq:A,B 2}). The dashed lines are the power series expansions around $g=0$ given by equations (\ref{eq:A power series}) and (\ref{eq:B power series}).
\label{fig:A B plot}}
\end{figure}

To investigate the solution to equation (\ref{eq:ODE}), we begin by studying the small $q$ behaviour. Expanding $\Gamma\left(q\right)$ at lowest order in a power series
\begin{equation}
\Gamma\left(q\right)=Aq^{\nu}+O\left(q^{\nu+1}\right)
\end{equation}
results in the equation
\begin{equation}
\left(4-\nu\right)\nu A^{2}q^{2\nu}+O\left(q^{2\nu+1}\right)=\frac{3\pi}{2}\frac{1}{\Lambda^{8}}q^{8}\,.
\end{equation}
Comparing the exponents of $q$ on the left and right-hand sides of this equation, we find that only $\nu=4$ can solve the equation at lowest order, however the factor of $\left(4-\nu\right)$ on the left-hand side ensures that $\nu=4$ will also fail to actually balance the left and right-hand sides. This observation forces the conclusion that the small $q$ expansion of $\Gamma\left(q\right)$ is not a power series. To obtain the actual small $q$ behaviour, the following heuristic argument can be made. Setting $\Gamma(q) = q^4u(q)$, equation (\ref{eq:ODE}) can be written as an ODE for $u(q)$
\begin{equation}
q^{2}u\frac{d^{2}u}{dq^{2}}+\frac{5}{2}q\frac{d\left(u^{2}\right)}{dq}=-\frac{3\pi}{2}\frac{1}{\Lambda^{8}}\,.
\end{equation}
Neglecting the second derivative term, this equation is exactly solvable, having solution
\begin{equation}
u\left(q\right)=\sqrt{A-\frac{3\pi}{5}\frac{1}{\Lambda^{8}}\ln\left(q\right)}
\end{equation}
where $A$ is some constant. Accordingly, we might expect the small $q$ behaviour of $\Gamma(q)$ to be logarithmically corrected by a term proportional to $\sqrt{\ln(q)}$ and indeed, without neglecting any terms, one can find upon substitution that for small $q$, equation (\ref{eq:ODE}) is solved by
\begin{multline}
\Gamma\left(q\right)=\left(\frac{q}{\Lambda}\right)^{4}\sqrt{\frac{3\pi}{4}\left[A-\ln\left(\frac{q}{\Lambda}\right)\right]}\times\\
\times\left[1+O\left(\frac{\ln\left(\ln\left(\Lambda/q\right)\right)}{\ln\left(\Lambda/q\right)}\right)\right]
\label{eq:Gamma small q expnsn}
\end{multline}
where $A$ is a constant determined by the boundary condition at $q=\Lambda$. For example, requiring that this equation match with equation (\ref{eq:large q BC}) gives
\begin{equation}
A=\frac{1}{8}+\frac{2}{3\pi}g^{2}+\frac{g}{\sqrt{6\pi}}\sqrt{1+\frac{8}{3\pi}g^{2}}\,.
\label{eq:A alone}
\end{equation}
Since equation (\ref{eq:Gamma small q expnsn}) is established around $q=0$ while equation (\ref{eq:large q BC}) imposes a boundary condition far from $q=0$, there is little reason to trust this approximation as we move away from $q=0$. To obtain reliable results across the entire domain $q\in\left[0,\Lambda\right]$, we can also expand the solution to equation (\ref{eq:ODE}) around $q=\Lambda$ and stitch the expansions together. To this end, we propose a type of two-point expansion that uses the known analytic behaviour around the two end points \cite{PadeApproximantsBook1996}
\begin{equation}
\Gamma\left(q\right)\simeq\left(\frac{q}{\Lambda}\right)^{4}\sqrt{\frac{3\pi}{4}\left[A-\ln\left(\frac{q}{\Lambda}\right)\right]}\left[1+B\frac{q}{\Lambda}\right]
\label{eq:ODE sol}
\end{equation}
where $B$ is another constant. For $q$ close to 0, this modification has no effect while for $q$ close to $\Lambda$, $B$ can be chosen to ensure that $\Gamma\left(q\right)$ be correct up to zeroth order. In particular, subbing this expression into the boundary condition given by equation (\ref{eq:large q BC}) gives
\begin{equation}
\frac{16}{\sqrt{3\pi}}g\sqrt{A}=\left(8A-1\right)+\left(10A-1\right)B\,,
\label{eq:A,B 1}
\end{equation}
while subbing it into equation (\ref{eq:ODE}) and expanding around $q=\Lambda$ implies that at lowest order, we must have
\begin{equation}
8A=\left(1+8A+B+12AB-20A^{2}B\right)\left(1+B\right)\,.
\label{eq:A,B 2}
\end{equation}
This pair of equations determines the constants $A$ and $B$. It is worth mentioning that the equations have multiple real solutions but only one of these solutions actually generates an approximation for $\Gamma\left(q\right)$ which approximately satisfies the original integral equation given by equation (\ref{eq:cont integ equ}). This useful solution is plotted in Fig.~\ref{fig:A B plot} as a function of $g$. Unfortunately, this single solution ceases to exist for values of $g$ greater than roughly 0.912 however since we are interested in small $g$, this poses little difficulty to us in practice. Furthermore, as $g$ grows, it turns out that using equation (\ref{eq:Gamma small q expnsn}) with equation (\ref{eq:A alone}) becomes an increasingly good approximation. Considering higher order corrections is also an option though this results in little gain at the cost of significantly increased complexity. 

\begin{figure}[t]
\includegraphics[width=\columnwidth]{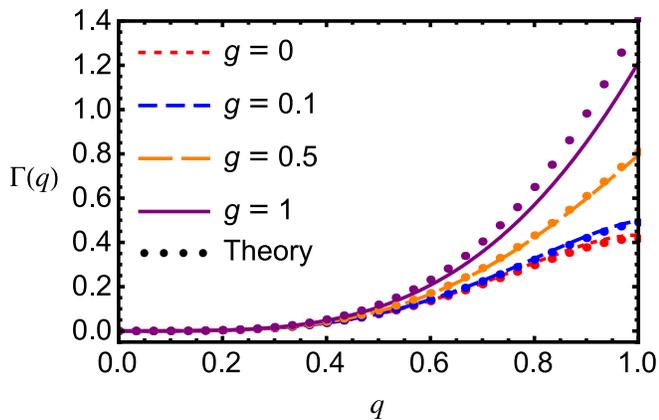}%
\caption{Comparison between numerical solutions of equation~(\ref{eq:ODE}) (solid and dashed lines) and the theory predicted by equation~(\ref{eq:ODE sol}) with $A(g)$ and $B(g)$ given by equations~(\ref{eq:A power series}) and (\ref{eq:B power series}) (dots). For small $g$, the theory exhibits excellent agreement with the numerical solutions.
\label{fig:ODE Sol}}
\end{figure}

It is not difficult to numerically evaluate equations (\ref{eq:A,B 1}) and (\ref{eq:A,B 2}) for any particular value of $g$. Since we are primarily interested in the case where $g$ is small, however, it is more convenient to have a small $g$ expansion for $A$ and $B$. This also has the added advantage of avoiding the spurious solutions of equations (\ref{eq:A,B 1}) and (\ref{eq:A,B 2}). To this end, we expand
\begin{equation}
A(g) = \sum_{n=0}^\infty A_n g^n \qquad B(g) = \sum_{n=0}^\infty B_n g^n
\end{equation}
where the $A_n$ and $B_n$ are constants. Subbing these expressions into equations (\ref{eq:A,B 1}) and (\ref{eq:A,B 2}) and equating coefficients of $g$ order by order, we obtain up to $O\left(g^{4}\right)$
\begin{multline}
A(g) \simeq 0.137+0.336g+0.243g^{2}+\\
+0.112g^3+O\left(g^{4}\right) \, , \label{eq:A power series}
\end{multline}
\begin{multline}
B(g) \simeq -0.265+0.360g-0.395g^{2}+\\
+0.311g^3+O\left(g^{4}\right) \, . \label{eq:B power series}
\end{multline}
These approximations are plotted as the dashed lines in Fig.~\ref{fig:A B plot} and can be seen to be faithful for values of \mbox{$g\lesssim0.4$}. Fig.~\ref{fig:ODE Sol} shows a comparison between equation~(\ref{eq:ODE sol}) with $A(g)$ and $B(g)$ given by equations~(\ref{eq:A power series}) and (\ref{eq:B power series}) (dots) and direct numerical solutions of equation~(\ref{eq:ODE}) (solid and dashed lines). The numerical solutions were obtained with $\Lambda=1$ via a shooting method \cite{NumericalRecipes2007} in which numerical integration of equation~(\ref{eq:ODE}) starting at $q=\Lambda$ was performed with varying initial values of $\Gamma(q=\Lambda)$ until solutions were found with the desired boundary condition given by equation~(\ref{eq:small q BC}). As can be seen, for small $g$, equation~(\ref{eq:ODE sol}) with $A(g)$ and $B(g)$ given by equations~(\ref{eq:A power series}) and (\ref{eq:B power series}) is a superb approximation for the numerical solution though the quality decreases once $g$ reaches the size of unity. It is understood that this is because the series expansions for $A(g)$ and $B(g)$ are only valid for small $g$ and indeed, as stated above, for non-small values of $g$ a better approximation is simply equation~(\ref{eq:Gamma small q expnsn}) with $A(g)$ given by equation~(\ref{eq:A alone}). Finally, we note that solutions to equation~(\ref{eq:ODE}) with differing frequency cut-offs $\Lambda$ are self-similar, that is, if $\Gamma(q)$ is a solution for a given  maximum frequency $\Lambda$, then the scaled solution $\Gamma(aq)/a^4$ with $a>0$ will be a solution with maximum frequency $a\Lambda$. Accordingly, the validity of the comparison shown in Fig.~\ref{fig:ODE Sol} is not limited to the case where $\Lambda=1$ since all other cases can be directly obtained by merely scaling the results.

Returning to the discrete system we began with, we obtain an approximate solution for $\Gamma_{\left|\vec{n}\right|}$ is
\begin{equation}
\Gamma_{\left|\vec{n}\right|}\simeq\left(\Lambda L\right)^{4}\Gamma\left(q=\frac{n}{L}\right)\,,
\end{equation}
or explicitly
\begin{multline}
\Gamma_{\left|\vec{n}\right|}\simeq n^{4}\sqrt{\frac{3\pi}{4}\left[A-\ln\left(\frac{n}{n_{max}}\right)\right]}\times\\
\times\left[1+B\frac{n}{n_{max}}\right]\,,
\end{multline}
where $n_{max}=\Lambda L$ denotes a discrete upper-frequency cut-off.

\bibliography{bibFile}

\end{document}